\documentclass[prl,reprint,showpacs,floatfix, superscriptaddress]{revtex4-1}
\usepackage[pdftex]{graphicx} 
\usepackage{dcolumn} 
\usepackage{bm} 
\usepackage{amssymb} 
\usepackage{placeins}
\usepackage{amsmath} 
\usepackage{footmisc}
\usepackage[latin1]{inputenc}
\usepackage{tikz}
\usepackage{footmisc}
\usetikzlibrary{shapes,arrows}
\usepackage{hyperref}

\begin{document}
\title{The $GW$ self-screening error and its correction using a local density functional}
\date{\today}
\author{J.\ Wetherell}
\affiliation{Department of Physics, University of York, Heslington, York YO10 5DD, United Kingdom}
\affiliation{European Theoretical Spectroscopy Facility}
\author{M.\ J.\ P.\ Hodgson}
\affiliation{Max-Planck-Institut f\"ur Mikrostrukturphysik, Weinberg 2, D-06120 Halle, Germany}
\affiliation{European Theoretical Spectroscopy Facility}
\author{R.\ W.\ Godby}
\affiliation{Department of Physics, University of York, Heslington, York YO10 5DD, United Kingdom}
\affiliation{European Theoretical Spectroscopy Facility}

\begin{abstract} 
The self-screening error in electronic structure theory is the part of the self-interaction error that would remain within the $GW$ approximation if the exact dynamically screened Coulomb interaction, $W$, were used, causing each electron to artificially screen its own presence. This introduces error into the electron density and ionization potential. We propose a simple, computationally efficient correction to $GW$ calculations in the form of a local density functional, obtained using a series of finite training systems; in tests, this eliminates the self-screening errors in the electron density and ionization potential.
\end{abstract}
\maketitle

The $GW$ approximation \cite{PhysRev.139.A796,1998RPPh...61..237A} within many-body perturbation theory is widely used to compute the self-energy in many-electron systems \cite{PhysRevB.81.085103, PhysRevB.81.085102, PhysRevLett.85.5611, PhysRevLett.110.146403, doi:10.1021/acs.jctc.5b01238, doi:10.1063/1.3089567, doi:10.1021/acs.jctc.6b00774}, but has limitations. Higher-order terms beyond $GW$ (known as vertex corrections) include a ``self-screening'' correction \cite{doi:10.1063/1.3249965} related to the self-interaction familiar in density-functional theory (DFT) \cite{HK64,KS65}. Attempts to correct the entirety of the self-interaction error via explicit vertex corrections have proved challenging \cite{PhysRevB.76.155106, 1402-4896-86-6-065301}. We adopt a physically more direct approach to the self-screening correction. 

A $GW$ calculation normally takes a DFT calculation as its starting point. The goal is to improve the quantities calculated from DFT, such as the ionization potential (IP) \cite{doi:10.1021/acs.jctc.6b00774}. However, the self-screening error is known to have an adverse effect on the IP. Furthermore, the effect of the $GW$ procedure on the electron density is not widely understood. We focus on the computation of the electron density from the Green's function \cite{Jack} and IP. We use the space-time method to solve Hedin's equations \cite{PhysRevLett.74.1827}. Thus the $GW$ method can be used in several flavors: one-shot ($G_{0}W_{0}$), semi self-consistency ($GW_0$) and full self-consistency ($GW$) \cite{doi:10.1063/1.3089567}. We identify the self-screening error inherent in all $GW$ calculations via the use of an effective potential. We then propose a simple and computationally inexpensive correction term that is a local potential added to the self energy and applicable to any $GW$ calculation. Finally we test our self-screening correction (ssc) by comparing the $GW+$ssc electron density and IP to the exact quantities from systems of few electrons in 1D where the many-electron Schr\"odinger equation (SE) can be solved exactly. For these model systems we find that the spurious effects of the self-screening error on the density and the IP are removed. 
 
Within $GW$ the screened interaction $W$ amounts to dynamically adjusting the strength of the bare Coulomb interaction between electrons. One merit of the exchange operator of Hartree-Fock theory is that it exactly corrects the self-interaction error introduced by the Hartree potential. If the Coulomb interaction in the Hartree-Fock exchange operator were screened using the exact irreducible polarizability $P$, Hartree-Fock's self-interaction correction would be improperly reduced, so that part of the self-interaction error -- the self-screening error -- remains uncorrected. It may be thought of as each electron artificially screening its own presence. It follows that the self-screening error is largest when screening, and therefore correlation, is strong \cite{doi:10.1063/1.3249965}.

The only source of error in a $GW$ calculation of the hydrogen atom (H) is self-screening because H is a one-electron system, and the RPA screening is exact for one electron (or strongly localized electrons) as no electron-hole interactions are present. This is apparent if one looks at the correlation part of the self-energy for this system, which should be zero. Instead, it consists of the spurious self interaction \cite{PhysRevA.75.032505}. 

\begin{figure}[htbp] 
\centering
\includegraphics[scale=0.55]{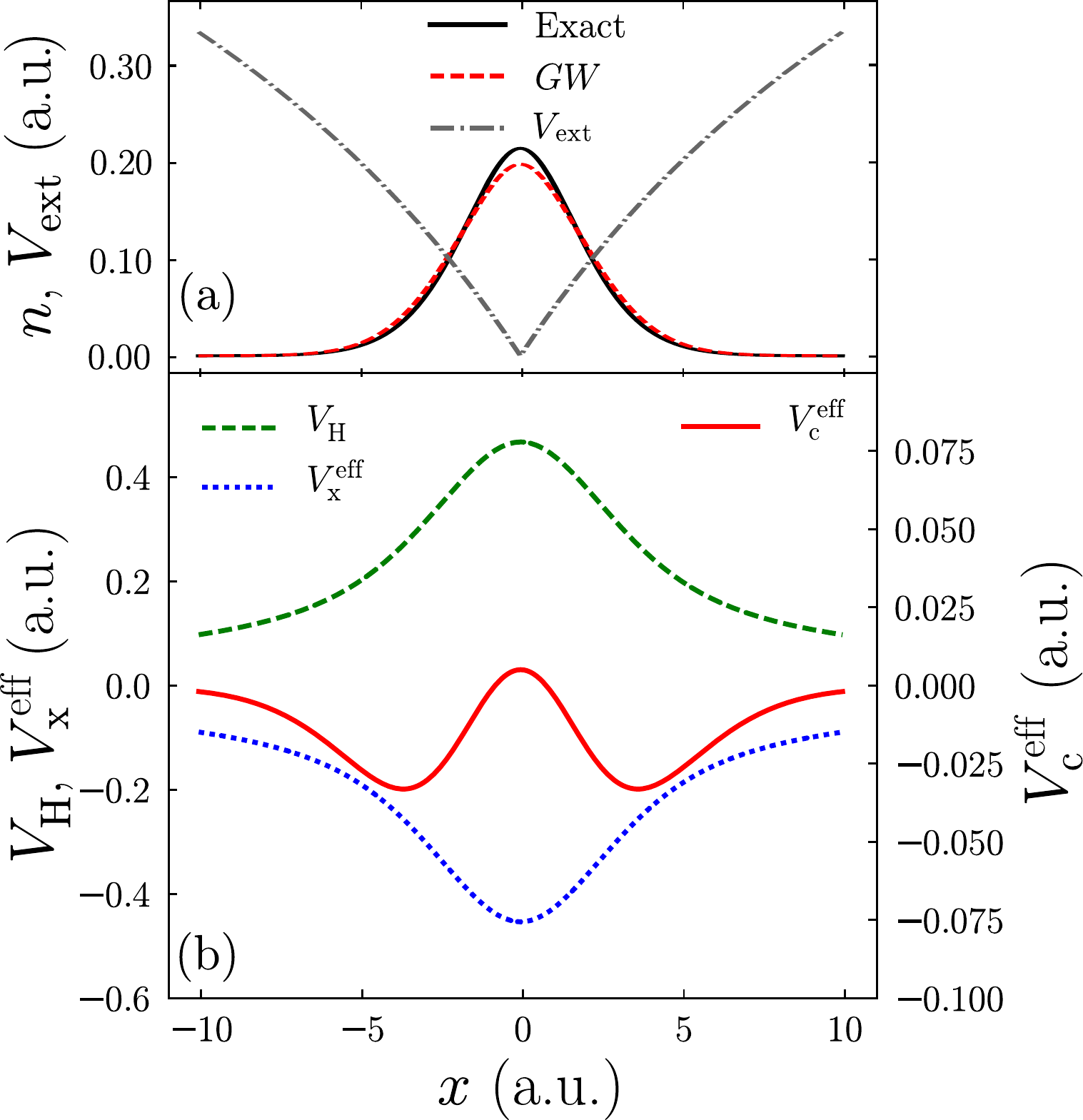} 
\caption{$GW$ self-screening error for a one-electron atom. (a) The exact density, fully self-consistent $GW$ density, and the external potential. The $GW$ density is more diffuse as the electron screens itself; the decay rate of the density towards the edge of the system is wrong. (b) The self-consistent $GW$ Hartree and effective exchange potentials (left scale), and effective correlation potential (right scale) (see Eq.~\ref{eff_pot_def}). The Hartree and effective exchange potentials cancel out, but there is a spurious non-zero effective correlation potential that results in the artificial spreading of the density away from the center of the system (a). This effective correlation potential is responsible for the entire self-screening error.}
\label{oec}
\end{figure}

We now employ a simple one-dimensional, one-electron model to investigate the self-energy of a $GW$ calculation. We compute all densities using our \texttt{iDEA} code \cite{PhysRevB.88.241102}, which determines the exact, fully-correlated, many-electron wavefunction in 1D, as well as our $GW$ densities. The electrons interact via the appropriately softened Coulomb repulsion $(|x-x'|+1)^{-1}$ \cite{PhysRevB.88.241102,PhysRevA.72.063411}. The electrons are treated as spinless to more closely approach the nature of exchange and correlation in systems of many electrons, hence each electron occupies its own distinct orbital \footnote{Spinless electrons obey the Pauli principle and are treated as same-spin.}. First we model \textit{one} electron in a 1D atomic potential, $V_\mathrm{ext}=-1.0/(\alpha|x|+1)$ where $\alpha=0.05$. The potential loosely confines the electron, which clearly displays the adverse effect of the self-screening error on the $GW$ density and energy (described above); see Fig.~\ref{oec}(a). We observe the effect self-screening has on the density by comparing the $GW$ density to the exact. As in this one-electron system, the screening has been accurately described by the RPA; the self-screening error is the only error present in the system, thus allowing us to investigate the effect of self-screening on the correlation part of the self energy.

Figure~\ref{oec}(a) shows the $GW$ and exact electron densities. Compared to the exact, the $GW$ density is more diffuse and its exponential decay is incorrect far from the center. If these inaccuracies in the density are a result of the self-screening, the correlation part of the self-energy will be non-zero. 
First we split the self-energy ($\Sigma$) into its separate contributions: $\Sigma = V_\mathrm{H} + \Sigma_\mathrm{x} + \Sigma_\mathrm{c}$, the Hartree potential ($V_\mathrm{H}$), exchange term ($\Sigma_\mathrm{x}$), and the correlation part ($\Sigma_\mathrm{c}$). $\Sigma_\mathrm{x}$ and $\Sigma_\mathrm{c}$ are non-local, and $\Sigma_\mathrm{c}$ is energy dependent. To get a clear picture of what effect a particular part of the self-energy is having on our only occupied orbital ($\phi$), we define an effective local potential, akin to DFT:
\begin{equation}
V_\mathrm{Hxc}^\mathrm{eff} \left(x\right) = \frac{1}{\phi \left(x\right)} \int \Sigma\left(x,x',\varepsilon \right) \phi \left(x'\right)dx',
\label{eff_pot_def}
\end{equation}
where $\varepsilon$ is the corresponding eigen-energy \footnote{\label{note1}As we use the space-time method we have access to the self-energy in the imaginary frequency domain, so for the purposes of Eq.~\ref{eff_pot_def} we approximate $\Sigma\left( \varepsilon \right) \approx \Sigma\left( \varepsilon_{f}\right)$ where $\varepsilon_{f}$ is the Fermi energy. To check this is a good approximation, once we have constructed the total effective potential, we solve the single-particle SE using $V_\mathrm{Hxc}^\mathrm{eff}$ to ensure this potential gives the same density as the $GW$ calculation.}. The three parts $V_\mathrm{H}$, $V^\mathrm{eff}_\mathrm{x}$, $V^\mathrm{eff}_\mathrm{c}$, defined in this way, may be examined separately.

Figure~\ref{oec}(b) shows the Hartree, effective exchange and correlation potentials for our one-electron atom. The $GW$ Hartree and effective exchange potentials completely cancel, as expected. However, there is a small correlation potential that is solely responsible for the error in the $GW$ density. We call this potential the self-screening potential $V_\mathrm{ss}$ which is present in all $GW$ calculations of any number of electrons ($N$). Examining the shape of $V_\mathrm{ss}$ we see this potential acts to draw the density away from the center of the system.

References~\onlinecite{Ren2012, doi:10.1063/1.4819399, doi:10.1080/00268976.2010.507227, doi:10.1080/00268976.2011.614282} note that a self-interaction error arises in the RPA total energy owing to the lack of a vertex in $P$. However, our preference is to focus on developing an effective vertex in the self-energy $\Sigma$, in order to retain the exact polarizability of a one-electron system.

Reference~\onlinecite{PhysRevB.85.035106} proposes to correct the self-screening error via an orbital- and spin-dependent screened interaction and is applicable to methods in which the Green's function is constructed from normalized single-particle orbitals. Our proposed self-screening correction, because it consists simply of a spatially local potential, is applicable to all flavors of $GW$. 

The potential $V_\mathrm{ssc}[n](x)$, that, when added to the $GW$ self-energy, $\Sigma_{GW}$, strives to yield a self-energy with self-screening removed: 
\begin{equation}
\Sigma_{GW+\textrm{ssc}} = \Sigma_{GW} + V_\mathrm{ssc}[n](x), \\
\label{define_sigma}
\end{equation}
where the density is obtained from the Green's function $G$.

\begin{figure}[htbp] 
\centering
\includegraphics[scale=0.6]{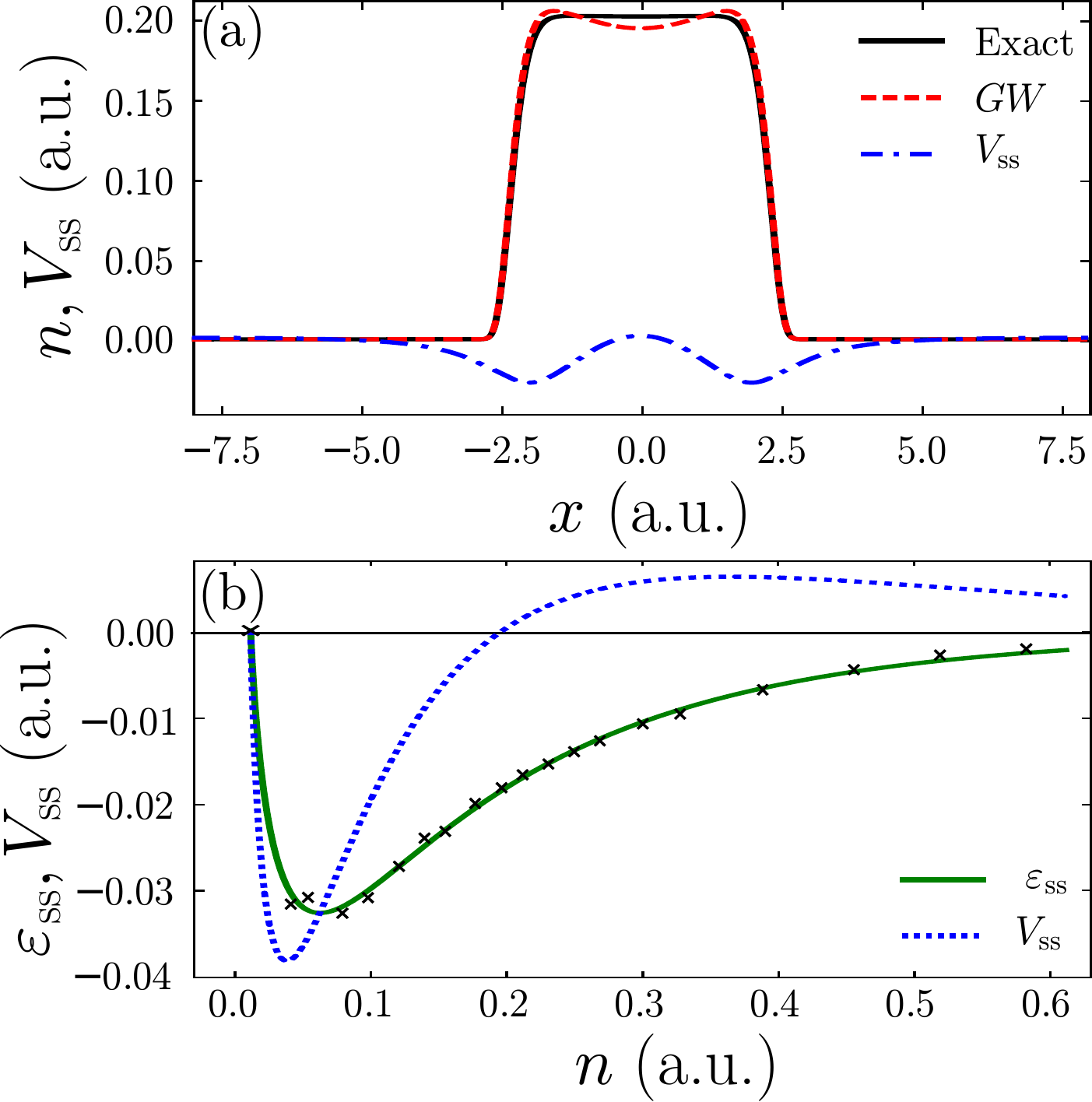} 
\caption{(a) An example one-electron finite, homogeneous density of height $n_0=0.22$, compared to the density produced by a self-consistent $GW$ calculation with the same external potential. The self-screening error causes the slab to curve as the electron screens itself. This is illustrated by the self-screening potential. (b) The crosses show the computed self-screening energy $\varepsilon_\mathrm{ss}$ per electron of each of the slab systems, with a cross added for $ \varepsilon_\mathrm{ss}(n=0)=0$. A fit for $\varepsilon_\mathrm{ss}$ is applied to these points. The corresponding self-screening potential $V_\mathrm{ss}$ is computed as the functional derivative of $\varepsilon_\mathrm{ss}$ (Eq.~\eqref{Vss}).}
\label{slabs}
\end{figure}

We construct a local density approximation for $V_\mathrm{ssc}[n](x)$. To do so we choose a set of finite, centrally homogeneous one-electron `density slabs', as used in Ref.~\onlinecite{PhysRevB.94.205134} to construct local-density approximations to the overall exchange-correlation functional of DFT 
\footnote{These `density slabs' are systems with uniform density within a finite region, resembling the homogeneous electron gas, and decaying to zero outside this region. Local approximations may therefore be constructed from results for a set of such slabs spanning a range of densities.}. 
We use one-electron slabs because, as we have established, in a one-electron system the self-screening is the only source of error in a $GW$ calculation. The density of each slab is chosen to be $n_0e^{- mx^{12}}+10^{-4} \cdot e^{-0.007 \left | x \right |}$, where $n_0$ is the height of the slab, and $m$ follows from normalization; see, e.g., Fig.~\ref{slabs}(a). Our set of slabs have a range of plateau densities $ 0.03 \leq n_0 \leq 0.58 $.  For each of these densities we apply the single orbital approximation (SOA) \cite{PhysRevB.90.241107,Hessler99}, which is exact for one electron, to obtain the external potential that defines this slab density. We then use our set of external potentials to calculate the corresponding exact total energy ($E$) for each slab density in turn via the single-particle SE. 

Next we perform a fully-self consistent $GW$ calculation for each slab system using the corresponding external potential to obtain the total $GW$ energy ($E_{GW}$). We choose to calculate $E_{GW}$ via the effective potential experienced by the single electron using Eq.~\eqref{eff_pot_def} -- note this is not the only means of calculating $E_{GW}$. To calculate $E_{GW}$, we construct an effective potential ($V^\mathrm{eff} = V_\mathrm{ext} + V^\mathrm{eff}_\mathrm{Hxc}$). We then solve the single-particle SE to find the lowest eigenvalue of this one-electron system, which is $E_{GW}$ \footnotemark[2]. Finally, we define the self-screening energy per electron as $\varepsilon_\mathrm{ss} = E_{GW} - E$.

Figure~\ref{slabs}(a) shows an example of one of our slabs with height $n_0=0.22$. The $GW$ density shows the effect of the self screening, and hence is not homogeneous in the central region whereas the exact is. Figure~\ref{slabs}(a) also shows the effective self-screening potential. Figure~\ref{slabs}(b) shows the self-screening energy as a function of the density $\varepsilon_\mathrm{ss}(n)$ for our whole range of slab systems (crosses). We require that $\varepsilon_\mathrm{ss}(n)$ must be zero when $n=0$ as there is no self-interaction, and therefore no self-screening error. We then apply a fit to this data yielding a functional form of the self-screening energy per electron:
\begin{equation}
\varepsilon_\mathrm{ss}(n(x)) = -a n(x) e^{-b n(x)^{c}},
\label{functiona1}
\end{equation}
where $a=4.09268$, b=$9.20609$ and c=$0.53652$.

Next we compute the local functional derivative, as follows \footnote{The total self-screening energy, $E_\mathrm{ss}$, is approximated to depend on the self-screening energy per electron as follows: $E_\mathrm{ss} = \int n\left(x\right)\epsilon_\mathrm{ss}\left(n\right)dx$}
\begin{equation}
V_\mathrm{ss}(n) = \varepsilon_\mathrm{ss}(n) + n\frac{d \varepsilon_\mathrm{ss}}{dn},
\label{Vss}
\end{equation}
in order to determine the local self-screening potential functional $V_\mathrm{ss}(n)$; shown in Fig.~\ref{slabs}(b). It follows that $V_\mathrm{ssc}[n](x) \approx -V_\mathrm{ss}(n(x))$,
in order for our correcting potential $V_\mathrm{ssc}$ to cancel the spurious self-screening of the electrons. Thus, our final local-density functional for correcting the $GW$ self-screening error is
\begin{equation}
V_\mathrm{ssc}(n(x)) = a n e^{-b  n^{c}}\left(2- b c n^{c}\right),
\label{functiona2}
\end{equation}
where $a$, $b$ and $c$ are given above. When we apply the $GW$ calculation with our local self-screening correction ($GW$+ssc) method to the training slabs we obtain the exact energy \footnote{But not the exact density, since the density of each slab system is extremely sensitive to the potential.}.

\begin{figure}[htbp] 
\centering
\includegraphics[scale=0.6]{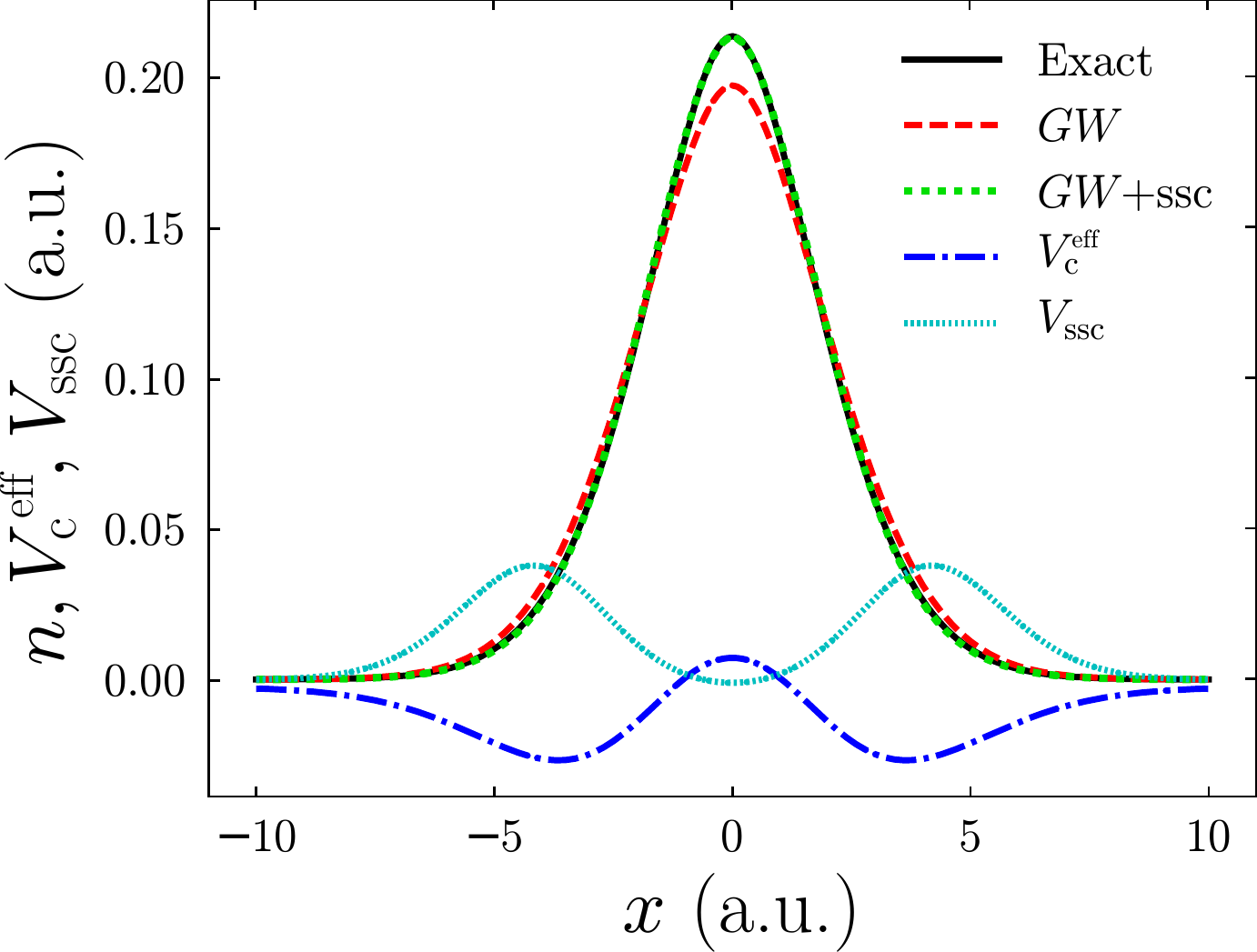} 
\caption{Applying our self-screening correction to the one-electron atom of Fig.~\ref{oec}. The $GW$ density is broadened relative to the exact due to the self-screening error, which is the only error present in the one-electron system. The self-screening-corrected $GW$ density is in excellent agreement with the exact, thus demonstrating the success of our functional. We can see that our self-screening correction potential successfully acts to cancel the self-screening potential present in the self-consistent $GW$ calculation, this also has the effect of correcting the decay of the effective xc potential far from the finite system.}
\label{correct_1}
\end{figure}

We test the effectiveness of our self-screening correction (Eq.~\eqref{functiona2}) by employing it for $GW$ calculations of various flavors (including self-consistent $GW$) for the one-electron atom described above, where self-screening is the only source of error. Figure~\ref{correct_1} shows the same one-electron model system as in Fig.~\ref{oec}; now the fully self-consistent $GW$+ssc is also shown. The $GW$+ssc density is in excellent agreement with the exact density, with the peak height and decay matching. (We show below that the IP predicted by the $GW$+ssc is also very accurate.) Figure~\ref{correct_1} also shows our local self-screening correction potential $V_\mathrm{ssc}$ and the $GW$ effective correlation potential $V^\mathrm{eff}_\mathrm{c} = V_\mathrm{ss}$; they cancel out very well thus removing the self-screening error, hence showing the success of correcting the self-screening error with a local potential. Here we only show the density calculated from self-consistent $GW$. However, we also find that the our self-screening correction is equally successful when applied to all of our flavors of $GW$ for this system.

\begin{figure}[htbp] 
\centering
  \includegraphics[scale=0.6]{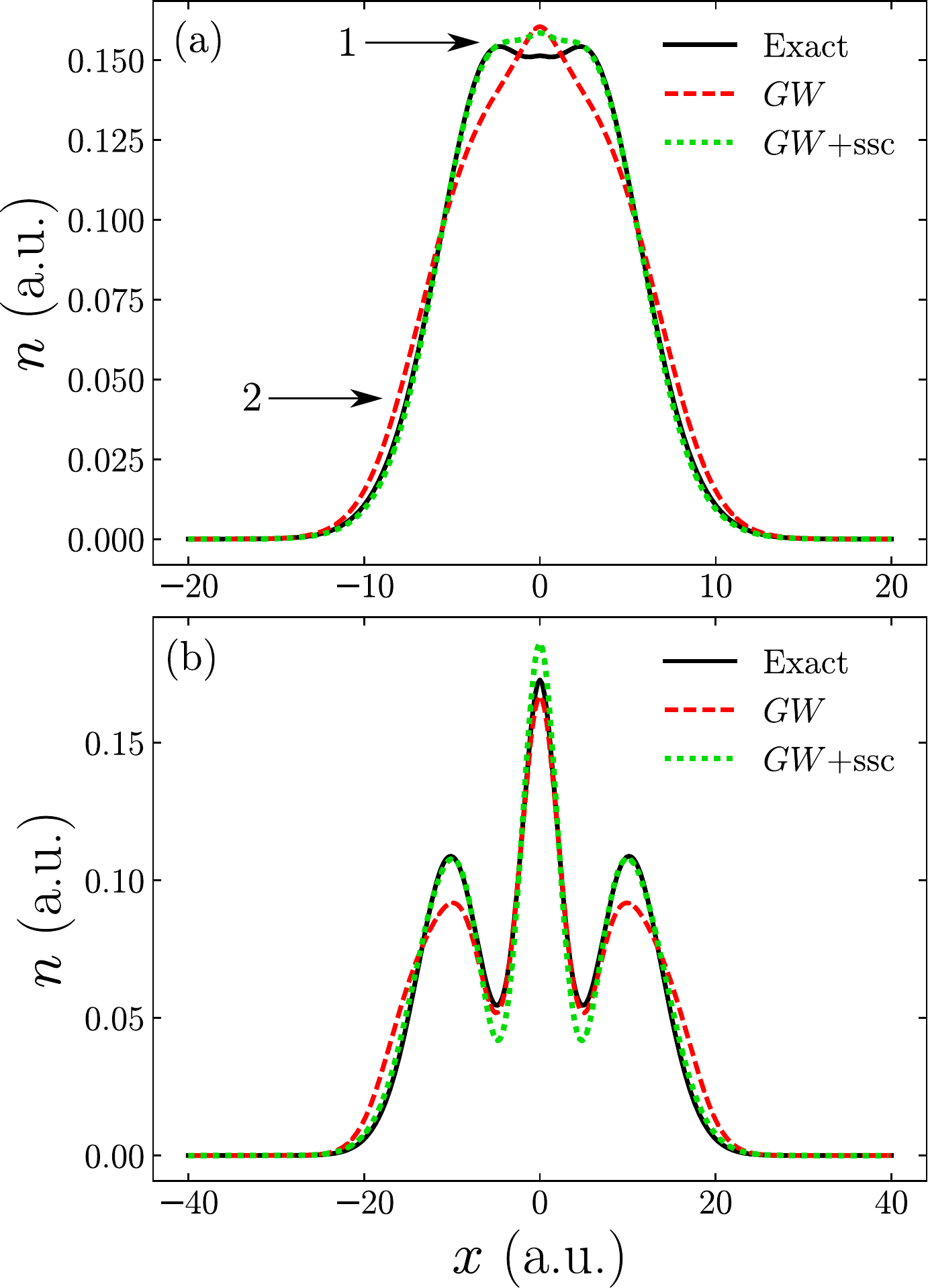} 
\caption{(a) Applying our self-screening correction to the two-electron atom. The $GW$ density is again dispersed compared to the exact. The self-screening corrected $GW$ density is in excellent agreement with the exact in region 2, where the electrons are strongly localized as the HOMO orbital is dominant in this region, and so the RPA gives a good account of the screening. Furthermore, the decay of the density is corrected, thus improving the IP of the system. In region 1 the density is less accurate owing to the delocalization of the electrons leading to electron-hole interactions present that are neglected in the RPA. Vertex correction to the polerizablity would be required to correct the density in this region \cite{PhysRev.84.1232}. (b) Applying our self-screening correction to the three-electron atom. The $GW$ density is once again diffuse compared to the exact. The self-screening corrected $GW$ density is in better agreement with the exact, it corrects the decay rate  and overall shape of the density, again improving the IP of the system. The height of the central peak remains incorrect, again due to the electron-hole interactions in this region neglected by the RPA.}
\label{correct_23}
\end{figure}

We also investigate two- and three- electron atoms using the same form of external potential as the one-electron atom (above), with $\alpha=0.05$ for two electrons and $\alpha=0.02$ for three electrons. Figure~\ref{correct_23}(a) shows that our self-screening correction significantly improves the $GW$ density for the two-electron atom; it corrects the decay rate of the density, thus improving the predicted IP (see Table~\ref{IPs}). The density maintains the incorrect central feature due to the electron-hole interaction neglected by the RPA in this delocalized region, which could in-principle be corrected by vertex corrections to $P$ \cite{PhysRev.84.1232}.

Figure~\ref{correct_23}(b) shows the effectiveness of our self-screening correction for the three-electron atom. Again, we find the self-screening-corrected $GW$ density is in better agreement with the exact; it corrects the decay rate of the density far from the center of the system, again improving the predicted IP (see Table \ref{IPs}), and the overall shape of the density. The height of the central peak remains incorrect, which again suggests a failing of the approximation to the polarizability $P$ and \textit{not} the presence of self screening. 

This ssc also improves $G_{0}W_{0}$ (one-shot) calculations. In this case the ssc is applied in the same way as above: the local potential is added to the self-energy via Eq.~\ref{define_sigma}. When applied to $G_{0}W_{0}$ starting from a conventional LDA calculation, the density errors are reduced by $16-50$\% for these model systems.

Table~\ref{IPs} shows the IPs predicted by $GW$ via two different methods for all three of our atoms, with and without our self-screening correction. Our first method extracts the IP from the density, which in principle can be done by determining the decay rate of the density far from the center of the system: $\lim_{\left | x \right | \rightarrow \infty} I = \frac{1}{4}\left( \frac{\partial \ln(n)}{\partial x}\right )^2$. In practice, we find it less computationally onerous to determine the exact KS potential corresponding to the $GW$ density for each atom in turn using the algorithm of Ref.~\onlinecite{PhysRevB.88.241102} and obtain the highest occupied KS eigenvalue which is the negative of the IP. When calculated in this way, the $GW+$ssc IP is strikingly accurate. (Similar results are obtained for non-self-consistent versions of $GW$.) We expect our ssc to similarly correct the IP for any $N$-electron system as localization becomes absolute far from the center of the system. Hence, in this region, the HOMO orbital is dominant and the RPA gives a good account of the screening. Therefore, the decay rate of the $GW+$ssc density is very accurate and thus so is the IP. This can be seen in Fig.~\ref{correct_23} (region 2 in (a)). Second, we calculate the IP via the quasiparticle (QP) energies. In contrast to the HOMO energy, these QP energies are affected by the electron-hole interactions present in the two- and three-electron atoms, hence the predicted IPs are not as accurate relative to extracting the IP from the density, but are generally improved by the ssc; see Table~\ref{IPs}.

\begin{table}[htb]
\caption{The IPs predicted by $GW$, and $GW$ with our self-screening correction, against the exact for the one, two and three electron atoms. IPs extracted via the KS potential and QP energies are shown.}
\label{IPs}
\resizebox{\columnwidth}{!}{
\begin{ruledtabular}
\begin{tabular}{c*{5}{>{$}c<{$}}}
       &  N  & GW & GW\rm{+ssc} & \rm{Exact} \\
  \hline
       &  1  & 0.908 & 0.900 & 0.900 \\
   KS  &  2  & 0.624 & 0.610 & 0.611  \\
       &  3  & 0.662 & 0.641 & 0.642 \\ \hline
       &  1  & 0.908 & 0.900 & 0.900 \\
   QP  &  2  & 0.577 & 0.577 & 0.611  \\
       &  3  & 0.675 & 0.654 & 0.642 \\
\end{tabular}\end{ruledtabular}}\end{table}

In conclusion, we propose a simple self-screening correction which is a local potential added to the self-energy of any $GW$ calculation. The correcting potential is a local functional of the electron density. We find that the self-screening error is removed from our $GW$ calculations of various test systems when our correction is employed. The electron density is significantly improved for all systems studied. In one-electron systems, and regions of high localization in many-electron systems, the density is almost exact. Beyond our self-screening correction, electron-hole corrections to the screening would be required in the delocalized regions. Furthermore the IPs predicted by $GW$ are improved by our correction. The method we used in this Letter for deriving our self-screening correction can be performed in 3D.

We acknowledge funding from the York Centre for Quantum Technologies (YCQT). We thank Krister Karlsson, Lucia Reining, Phil Hasnip, Leopold Talirz, Mike Entwistle and James Quirk for fruitful discussions.

\bibliography{jw1294.bib}

\end{document}